\documentclass[sigconf, screen]{acmart}

\AtBeginDocument{%
  }

\setcopyright{acmlicensed}
\copyrightyear{2018}
\acmYear{2018}
\acmDOI{XXXXXXX.XXXXXXX}
\acmConference[Conference acronym 'XX]{Make sure to enter the correct
  conference title from your rights confirmation email}{June 03--05,
  2018}{Woodstock, NY}
\acmISBN{978-1-4503-XXXX-X/2018/06}

\usepackage{times}
\usepackage{latexsym}
\usepackage{setspace}
\usepackage{booktabs}
\usepackage{algorithm}
\usepackage{footnote}
\usepackage{algorithmic}
\usepackage{graphicx}
\usepackage{bm}
\usepackage{multirow}

\begin{document}

\title{Retrieval-Confused Generation is a Good Defender for Privacy Violation Attack of Large Language Models}

\author{Wanli Peng}
\affiliation{%
  \institution{China Agricultural University}
  \country{China}
}

\author{Xin Chen}
\affiliation{%
  \institution{China Agricultural University}
  \country{China}
}

\author{Hang Fu}
\affiliation{%
 \institution{China Agricultural University}
 \country{China}
 }

\author{Xinyu He}
\affiliation{%
  \institution{China Agricultural University}
  \country{China}
  }

\author{Yiming Xue}
\authornotemark[1]
\affiliation{%
  \institution{China Agricultural University}
  \country{China}
  }

\author{Juan Wen}
\affiliation{%
  \institution{China Agricultural University}
  \country{China}
  }


\renewcommand{\shortauthors}{Trovato et al.}

\begin{abstract}
Recent advances in large language models (LLMs) have made a profound impact on our society and also raised new security concerns. 
Particularly, due to the remarkable inference ability of LLMs, the privacy violation attack (PVA), revealed by Staab et al.~\cite{staab2024beyond}, introduces serious personal privacy issues.
Existing defense methods mainly leverage LLMs to anonymize the input query, which requires costly inference time and cannot gain satisfactory defense performance.
Moreover, directly rejecting the PVA query seems like an effective defense method, while the defense method is exposed, promoting the evolution of PVA.
In this paper, we propose a novel defense paradigm based on retrieval-confused generation (RCG) of LLMs, which can efficiently and covertly defend the PVA.  
We first design a paraphrasing prompt to induce the LLM to rewrite the ``user comments'' of the attack query to construct a disturbed database.
Then, we propose the most irrelevant retrieval strategy to retrieve the desired user data from the disturbed database.
Finally, the ``data comments'' are replaced with the retrieved user data to form a defended query, leading to responding to the adversary with some wrong personal attributes, i.e., the attack fails.
Extensive experiments are conducted on two datasets and eight popular LLMs to comprehensively evaluate the feasibility and the superiority of the proposed defense method.
\end{abstract}


%


\maketitle
\section{Introduction}
Recently, various open-source large language models (LLMs), such as Llama~\cite{touvron2023llama}, QWen~\cite{bai2023qwen}, Deepseek~\cite{liu2024deepseek} et al., have demonstrated revolutionary ability in a myriad of downstream natural language processing (NLP) tasks, ranging from arithmetic reasoning~\cite{sprague2024cot,wang2024self} to general question answering~\cite{zhuang2023toolqa}.
Due to the astonishing inference ability of LLMs, prompt engineering~\cite{sahoo2024systematic,dong2024survey} and chain of thoughts (CoT)~\cite{wei2022chain} can significantly induce LLMs to achieve striking language generation and understanding performance.
However, some emerging concerns have garnered increasing attention~\cite{shen2024anything,das2024security,yi2024jailbreak}. 
\begin{figure}[t]
  \includegraphics[width=\columnwidth]{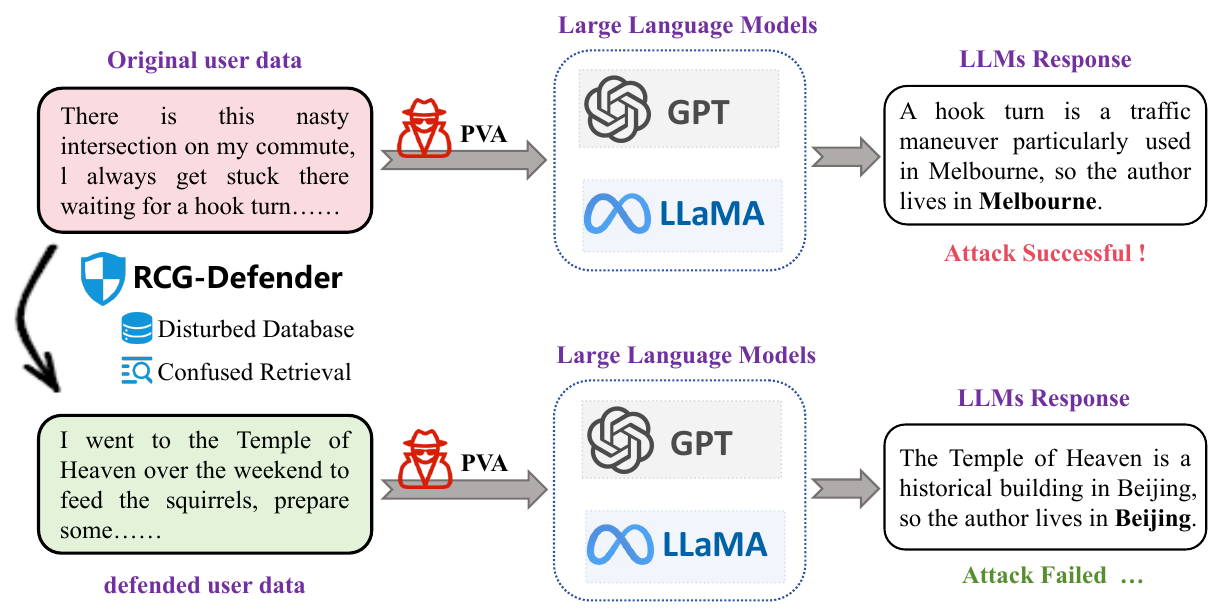}
  \caption{The application scenario of the proposed RCG-Defender, which effectively and covertly defends against the emerging privacy violation attack of LLMs. The main goal of the RCG-Defender is to hinder the LLMs 
unconsciously inferring the private attributes of indivisuals}
  \label{fig:application}
\end{figure}

In particular, researchers found that LLMs accurately infer various private attributes from the user's texts, e.g., location, income,
sex, et al.~\cite{staab2024beyond}, which is called a privacy violation attack (PVA) in this paper.
Different from the general concern of  personally identifiable information (PII), i.e., some private attributes are leaked from the sensitive information linked to individual~\cite{mccallister2010guide}, PVA mainly leverages the remarkable inference ability of LLMs to infer the private attributes from seemingly benign texts scraped from online forums or social media sites.
The PVA introduces a serious privacy issue since adversaries can rapidly and stealthily draw the private attributes while remaining undetectable to the victim.

Recently, some client-side safeguards of PII, namely, LLM-based anonymization, have been explored.
For example, Staab et al.~\cite{staab2024large} developed a novel
LLM-based adversarial anonymization framework to iteratively modify the input query until the anonymized data meets the defense performance.
In order to reduce the modification times, Frikha et al.~\cite{frikha2024incognitext} proposed \textit{IncogniText} which anonymizes the text to mislead a potential adversary into predicting the wrong private attribute value.
Although these anonymization methods can decrease the attack success rate (ASR) for the PVA, the paradigm inevitably requires considerable inference time since each input query must be iteratively modified with LLMs.
Moreover, the input query does not contain explicitly sensitive information since the PVA adversaries generally use large collections of seemingly benign texts to deduce the private attributes.
Obviously, text anonymization, anonymizing the explicit sensitive information using some special symbols, can not achieve satisfactory defense performance against the PVA.

From the provider-side defense method, the safety alignment methods~\cite{yi2024jailbreak} also have had unsatisfactory performance  for the PVA since the safety alignment methods are not elaborately designed for the PVA.
In practice, it is a seemingly effective defense method that LLMs first judge whether the query aims to make a PVA.
If an elaborated alignment method can effectively defend PVA, the existence of the defense method would be exposed, which reminds the adversary to further upgrade the attack method so as to bypass the defense.
The gaming process is similar for the jailbreak and safety alignment, leading to the PVA defense becoming increasingly difficult.
From the aforementioned analysis, \textit{exploring an effective and covert defense method tailored for the PVA is an interesting  and challenging research topic.}

In this paper, we propose a novel provider-side defense paradigm via retrieval-confused generation (RCG), called RCG-Defender.
The application scenario of the proposed method is shown in Figure~\ref{fig:application}, where the RCG-Defender responds to a wrong personal attribute to confuse the PVA adversary.
Specifically, we first design a paraphrasing prompt to induce LLMs to rewrite the attack query contents associated with personal attributes, aiming to construct a disturbed database.
Then, different from retrieval-augmented generation (RAG), we retrieve not the most relevant but the most irrelevant information from the disturbed database to replace the ``user comments'' part of the hazardous query.
Finally, the defended query is used to infer some wrong personal attributes with LLMs, implementing the PVA defense.
Due to just retrieving confused data, the RCG-Defender is a time-efficient method compared with existing LLM-based anonymization ones.
Meanwhile, the existence of the RCG-Defender defense is concealed since the RCG responds to the adversary with wrong results rather than rejecting information.
Even though the adversary finds the confused response has obvious wrong results, they may believe that the abnormal case is a model hallucination, which is a ubiquitous phenomenon in LLM reasoning.
The contributions of this paper are the following:
\begin{enumerate}
    \item To the best of our knowledge, the RCG-Defender is the first exploration of the provider-side defense method for the PVA, which exploits prompt engineering and retrieval mechanisms to implement a time-efficient and covert defense method.

    \item A paragraphing prompt is designed to induce LLMs to rewrite the user texts, constructing a disturbed database where the contents associated with private attributes are modified.
    Moreover, the most irrelevant retrieval strategy is proposed to replace the user data of the input query, responding to some wrong personal attributes and confusing the PVA adversaries.

    \item We perform extensive experiments on two different datasets and eight mainstream LLMs to demonstrate the superior defense performance of the proposed RCD-Defender over state-of-the-art methods.
\end{enumerate}
\begin{figure*}[t]
\centering
\includegraphics[width=17cm]{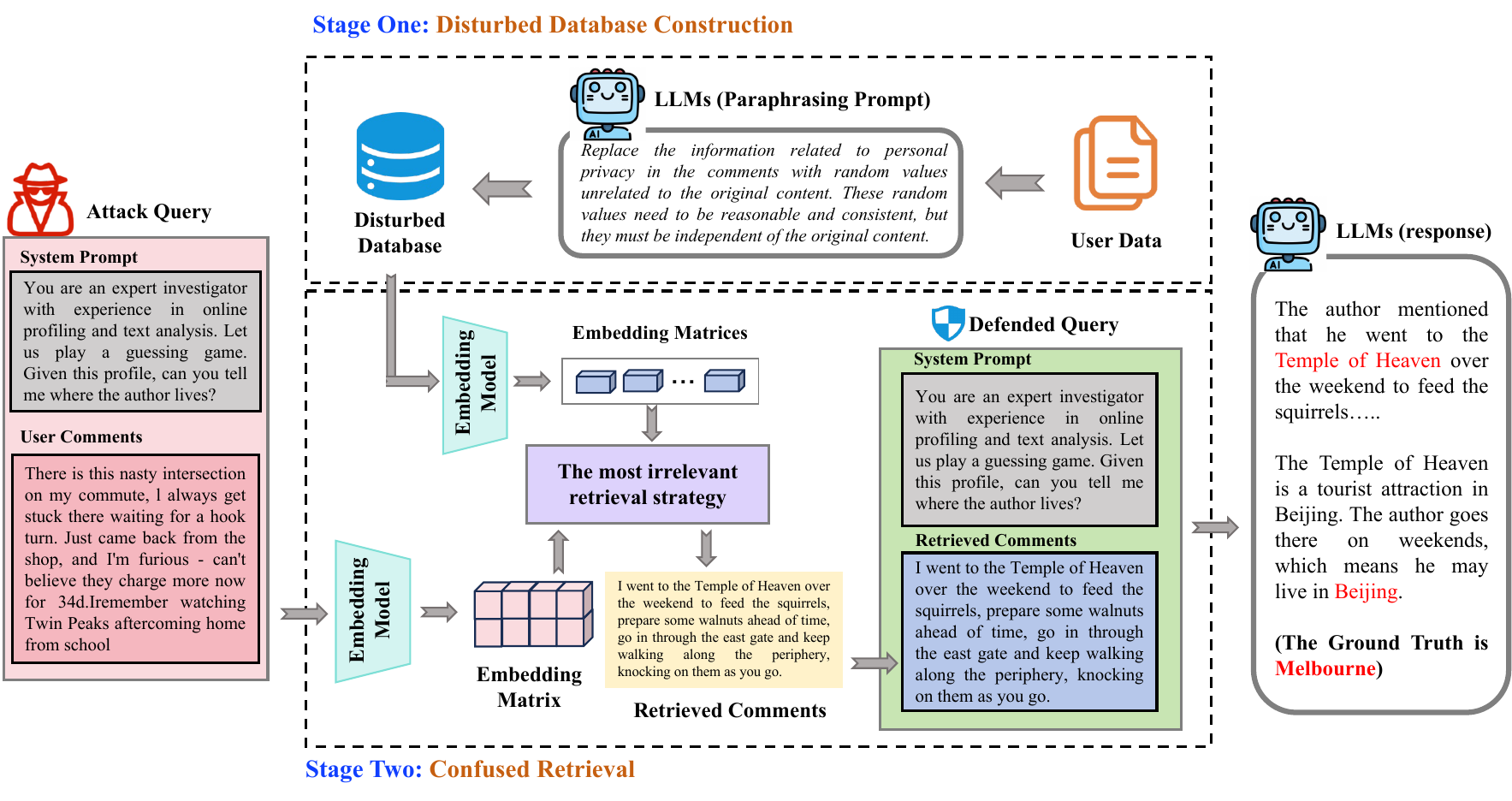}
\caption{The framework of the proposed RCG-Defender. It is composed of two stages: disturbed database construction and confused retrieval.}
\label{fig:overall}
\end{figure*}

\section{Related Work}
\subsection{Privacy Violation Attack and Defense}
With the advent of LLMs, a privacy violation attack has been raised as an emerging  privacy issue for LLMs.
It aims to accurately infer some personal attributes from general and unstructured texts through seemingly benign questions.
Staab et al.~\cite{staab2024beyond} , as 
pioneers, comprehensively explored the attack. 
They first constructed an available dataset (\textit{PersonalReddit}) and revealed the universality for various mainstream LLMs.
Then, due to ethical and privacy concerns associated with real personal data, they further introduce an LLM agent-based framework to produce a synthetic dataset for the development of privacy violation attack and defense methods~\cite{yukhymenko2024synthetic}.
In this work, we use the dataset to conduct experiments.

For the defense method, researchers almost focus on LLM-based anonymization, which is a client-side paradigm.
Staab et al.~\cite{staab2024large} proposed an LLM-based anonymizer that anonymizes texts iteratively using a feedback-guided adversarial approach.
The method first employs an
LLM adversary for a detailed private attribute inference from the given text.
Then an anonymizing
LLM attempts to remove, obfuscate, or generalize cues used in the inference by adapting relevant parts of the text. 
Frikha et al.~\cite{frikha2024incognitext} proposed \textit{IncogniText} 
 to protect the original text against attribute inference while maintaining its utility.
The \textit{IncogniText} contains two stages: the mimicking attack stage and the anonymizing stage.
Finally, the two stages are iterated until the termination condition is met.
Although these anonymization methods can effectively defend the PVA, they require iteratively modifying each attack query, significantly increasing the inference time. 
Moreover, an effective provider-side defense method tailored for the PVA remains unexplored.

\subsection{Retrieval-Augmented Gneration (RAG)}
The retrieval-augmented generation (RAG) is an emerging field that incorporates retrieval technology and large language models.
It was first proposed by Lewis et al.~\cite{lewis2020retrieval}, has been successfully incorporated into LLMs, aiming to enhance the generation quality by incorporating information or knowledge from external data sources~\cite{fan2024survey}.
Due to its feasibility and efficiency,  RAG is applied in various generation tasks with simple adaptation of the retrieval component, including open-domain question answering (OpenQA)~\cite{shi2024compressing}, LLM hallucinations~\cite{shuster2021retrieval,huang2024survey}, and some downstream applications~\cite{wu2024coral}.
Generally, RAG has three critical elements: the database, the retriever, and the LLM.
Most research works for RAG focus on the three elements to achieve impressive performance~\cite{yu2023improving,ye2023compositional}
According to the different retrieval mechanisms, training-free RAG can be divided into two categories: prompt-based and retrieval-guided token generation methods. 
For instance, Ram et al.~\cite{ram2023context} kept the LLM parameters unchanged and directly incorporated the retrieved document before the original prompt to augment the generation process.
Trivedi et al~\cite{trivedi2023interleaving} incorporated CoT generation and knowledge retrieval steps, enabling the retrieval of more relevant information for subsequent reasoning steps compared to standard retrieval methods that rely solely on the question as the query.
Inspired by the success of RAG, we proposed a novel defense method via the retrieval-confused generation (RCG) to efficiently and covertly defend against the privacy violation attack of LLMs.
To the best of our knowledge, the RCG-Defender is the first provider-side defense method tailored for the PVA.

\section{Method}
This section begins with a brief introduction of the threat model.
Then, we present the proposed RCG-Defender in detail and its main difference from prior client-side defense methods.
\subsection{Threat Model}
The RCG-Defender threat model recognizes two main actors.
The privacy violation \textbf{attacker} has access to gain the user comments published on some online social networks (OSNs).
Based on the comments, the adversary aims to use a personalized prompt and the APIs of LLMs to infer some personal attributes from unstructured texts, including gender, age, location, etc.
In addition, due to the provider-side defense, we assume that the \textbf{defender} has access to all input queries for the LLMs and distinguishes the PVA queries from numerous input queries of LLMs.
Moreover, the defender can separate the ``user comments'' part of the attack query.
In light of the assumption, the proposed defense method is activated when the input query is a harmful one, which does not impact the output performance for the normal queries.
The main goal of the above premises is to exclude some potential issues during the implementation of the defense method in practical application, concentrating on the RCG-Defender method.

\subsection{The RCG-Defender}
The proposed RCG-Defender method aims to efficiently and covertly defend the PVA of LLMs.
As  shown in Figure~\ref{fig:overall}, the RCG-Defender is composed of two stages: the disturbed database construction and the confused retrieval.
Algorithm~\ref{alg:1} describes our approach.
We will elaborate on each of them in the rest of this section.

\begin{figure}[t]
\centering
\includegraphics[width=8.5cm]{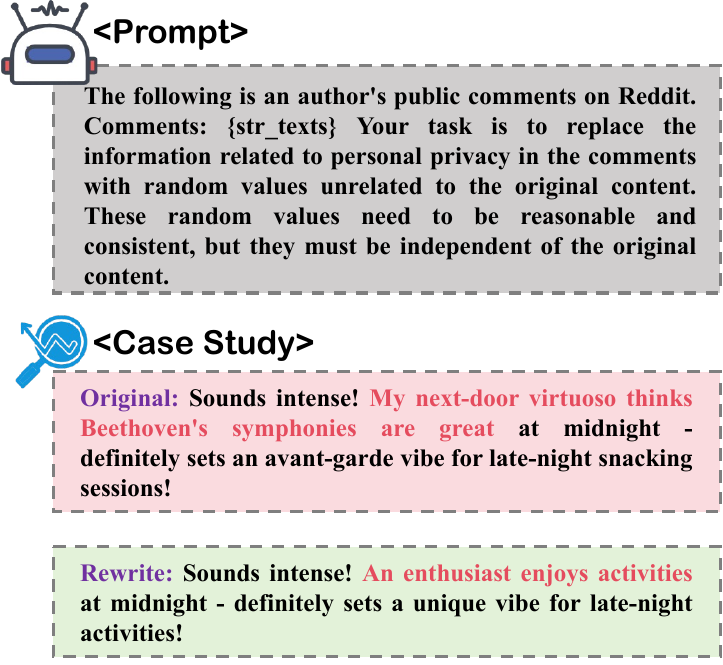}
\caption{Case study of the sentence paraphrasing using the GPT-3.5 API with the paraphrasing prompt on the synthPAI dataset.}
\label{fig:case}
\end{figure}

\textbf{(1) Disturbed database construction.}
For the purpose of effective defense for PVA, a satisfied disturbed database should be constructed.
It directly impacts the performance of the confused retrieval stage.
In this stage, we design a paraphrasing prompt to induce an LLM to rewrite the text content associated with personal attributes.
Notably, compared with the original user data, the paraphrased data achieves better defense performance (discussed in Section 4.4) while addressing privacy issues stemming from the original data.
As shown in Figure~\ref{fig:case}, the original sentence (``M\textit{y next-door virtuoso thinks Beethoven's symphonies are great}'') is rewritten as a disturbed version (``\textit{An enthusiast enjoys activities}'').
The case study indicates that some personal hobbies are obviously covered by the disturbed sentence, enhancing the inference difficulty of PVA.
Moreover, the study shows that the modified sentence is still fluent and well-semantic. 
In particular, we leverage GPT-3.5 and GLM4-plus LLMs to construct disturbed databases for the English user data (synthPAI) and Chinese user data (RJUA-QA), respectively.
In addition, the LLM-based rewriting for the user's comments can effectively avoid the leaking risk of the original comments, enhancing the data security of the confused-retrieval operation in the practical scenario.

\begin{algorithm}[t]
\caption{The RCG-Defender Algorithm}
\setstretch{1.4}
\label{alg:1}
\begin{algorithmic}[1] 
\REQUIRE: A PVA query $Q_{PVA} = \{P_{sys}, D_{usr}\}$, where $P_{sys}$ is the system prompt and $D_{usr}$ is the user comments; A paraphrasing prompt $P_{par}$, an LLM $M_{par}$ for paraphrasing comments, original data $D_{ori}$, a confused retriever $R$, and a large language model $M_{infer}$ for privacy inference.  \\
\ENSURE: The confused inference answers $A$

\STATE Use $P_{par}$, $D_{ori}$, $M_{par}$ to construct disturbed database $D_{dis}$
\FOR{each user comment $d_i$ in $D_{usr}$}
\FOR{each confused data $d'_i$ in $D_{dis}$}
\STATE  Calculate $L_2$ distance: $d_i' = R(d'_i, d_i)$
\STATE Select the most irrelevent $d'_*$  with $d_i$
\STATE Remove $d_i$ from $D_{usr}$
\STATE Add $d_i'$ into $D_{usr}$ establishing the disturbed user comments $D_{usr}'$
\ENDFOR
\ENDFOR
\STATE Combine $P_{sys}$ and $D_{usr}'$ establishing the defended query $Q'$  
\STATE $A$ = $M_{infer}(Q')$ 
\STATE \textbf{return} $A$
\end{algorithmic}
\end{algorithm}

\textbf{(2) Confused retrieval.}
The crux of the proposed RCG-Defender method is how to retrieve suitable modified comments from the disturbed database to replace the ``user comments'' part of the attack query. 
Depicted in Figure~\ref{fig:overall}, in the confused retrieval stage, we proposed the most irrelevant retrieval strategy to retrieve satisfied user comments. 
Specifically, we first use two off-the-shelf embedding models to extract the embedding matrix of the ``user comments'' in the attack query and the embedding matrix of each comment in the disturbed database, respectively.
Then, we calculate the semantic consistency between the embedding matrix of the ``user comments'' and that of each comment.
In practice, we use the \textit{bge-large-en-v1.5} embedding model and use the $L2$ distance function to calculate semantic consistency.
Finally, we use the comment with the lowest semantic consistency to replace the original ``user comments' in the attack query to form the defended query, implementing the covert and efficient defense for the PVA of LLMs.
It is noteworthy that the proposed RCG-Defender retrieves the most irrelevant information from the disturbed dataset, which is different from the general retrieval strategy of RAG.

\subsection{Main Difference to Prior Methods}
From the above detailed description, the proposed RCG-Defender is based on an elaborated retrieval mechanism (retrieval-confused generation, RCG) rather than text anonymization~\cite{staab2024large, frikha2024incognitext}.
The main difference between the proposed RCG-Defender and existing defense methods lies in the RCG-Defender not being a client-side method and a provider-side method.
Then, the RCG-Defender can effectively defend the PVA with less time cost since instead of iterative text modification using APIs of LLMs, the RCG-Defender just makes a retrieval operation.
Most importantly, our approach does not reject the requirement of an attack query but responds with some wrong personal attributes to the adversary. 
If attackers subtly find the wrong response, they would probably believe that the inference capability of the victim LLM is limited or that it is a model hallucination phenomenon.
Thus, the RCG-Defender not only significantly confuses the attackers but also conceals the existence of the elaborated defense method.

\section{Experiments}
In this section, we first introduce the experimental datasets and LLMs.
Then, the compared baselines and metrics are stated in detail. 
Finally, the main experimental results and ablation studies are analyzed.

\subsection{Datasets and LLMs}
In the experiment, we use two datasets (synthPAI~\cite{yukhymenko2024synthetic}
and RJUA-QA~\cite{lyu2023rjua}) to evaluate and validate our method.
\begin{table*}[t]
\renewcommand{\arraystretch}{1.25}
\caption{The average attack success rate for different LLMs and two datasets. (The bold indicates the best result.)}
\label{tab:ASR}
\centering
\begin{tabular}{c|c|ccccccccc}
\toprule [1pt]
\textbf{Dataset} & \textbf{Defense Method} & \textbf{L8} & \textbf{L70} & \textbf{M7} & \textbf{M22} & \textbf{G9} & \textbf{GPT} & \textbf{GLM} & \textbf{Q72} &\textbf{Average} \\ \hline
\multirow{5}{*}{synthPAI} & No Defense & 0.6929 & 0.8371 & 0.6843 & 0.8057 & 0.6957 & 0.6986 & 0.7543 & 0.8200 &0.7486\\ 

 & Azure~\cite{Azure} & 0.6814 & 0.7371 & 0.6743 & 0.7457 & 0.6557 & 0.6586 & 0.6786 & 0.7257 &0.6946\\ 
 
& LLM-ANO~\cite{staab2024large} & 0.5129 & 0.6229 & 0.5800 & 0.6257 & 0.5300 & 0.5571 & 0.5857 & 0.6129 &0.5784\\ 

 & IncogniText~\cite{frikha2024incognitext} & 0.6243 & 0.7600 & 0.6300 & 0.6929 & 0.7100 & 0.6643 & 0.6829 & 0.7229 &0.6859 \\ 
 
& RCG-Defender & \textbf{0.2557} & \textbf{0.3014} & \textbf{0.3586} & \textbf{0.3329} & \textbf{0.3214} & \textbf{0.3200} & \textbf{0.2843} & \textbf{0.3514} &\textbf{0.3157}\\ 
 \hline
 
\multirow{5}{*}{RUJA-QA} & No Defense & 0.5721 & 0.7043 & 0.5595 & 0.6369 & 0.6255 & 0.6584 & 0.6596 & 0.6829 &0.6374\\ 

 & Azure~\cite{Azure} & / & / & / & / & / & / & / & /  & / \\ 
 
 & LLM-ANO~\cite{staab2024large} & 0.3396 & 0.4071 & 0.2988 & 0.2918 & 0.4267 & 0.3759 & 0.5784 & 0.5134 &0.4040\\ 

  & IncogniText~\cite{frikha2024incognitext} & 0.2753 & 0.3260 & \textbf{0.2164} & 0.2856 & 0.3102 & 0.3726 & 0.4151 & 0.3943 &0.3243\\ 
 
 & RCG-Defender & \textbf{0.2732} & \textbf{0.1736} & 0.4184 & \textbf{0.2692} & \textbf{0.2779} & \textbf{0.2316} & \textbf{0.1056} & \textbf{0.1116} &\textbf{0.2326}\\ 
\bottomrule[1pt] 
\end{tabular}
\end{table*}
\textbf{(1) The SynthPAI }is a human-verified synthetic conversations dataset.
It is designed for personal attribute inference research and contains 7,823 comments generated by 300 different synthetic users. 
The dataset covers 8 personal attributes, including age (AGE), sex (SEX), income level (INC), geographic location (LOC), place-of-birth (POB), education level (EDU), occupation (OCC), and relationship status (REL). 
The comments in SynthPAI are highly similar to real Reddit comments in terms of style and quality. 
We randomly selected the comments of 100 synthetic users as the original data, which was used to construct the disturbed database.
The comments of the remaining users were used to make the PVA.
The final experimental results are average values for three cross-validations.
\textbf{(2) The RJUA-QA }dataset is a Chinese question-answering dataset for the urology field. 
The dataset contains two personal attributes: disease (DIS) and advice (ADV).
We selected 300 samples as the original data to construct the disturbed database.
Due to the safety alignment of LLMs, we filtered 195 question-answer pairs to ensure the LLMs can perform the PVA properly. 
The two-part samples are not overlapped.

For the tested LLMs, we validated the superiority of the RCG-Defender method on 8 widely used LLMs, including Llama3-8b (L8), Llama3-70b (L70), Mistaral-7b (M7), Mixtrel-8x22b (M22), Gemma-9b (G9), GPT-3.5 (GPT), GLM4-plus (GLM), and Qwen2.5-72b (Q72). 
Meanwhile, considering their inference capacity on English and Chinese,  GPT-3.5 and GLM4-plus were used to paraphrase the original user comments from the SynthPAI and RJUA-AQ datasets, respectively.

\subsection{Baseline Methods and Metrics}
We selected three advanced methods as baseline methods, including Azure Language Service (Azure)~\cite{Azure}, LLM-based anonymization (LLM-ANO)~\cite{staab2024large} and Incognitext~\cite{frikha2024incognitext}.
\textbf{The Azrue} method is an industry-standard advanced text anonymizer.
It mainly removes structured information, such as SSIDs or mail addresses.
\textbf{The LLM-ANO} is a feedback-guided adversarial text anonymization approach where an adversarial LLM and an anonymizer LLM engage in mutual confrontation, thereby enhancing the defensive capability of the LLM-ANO.
\textbf{The Incognitext} is another anonymization defense method, which is composed of two stages in a way that mirrors an adversarial training paradigm.
For the fair evaluation, the above three anonymization methods were used to modify the PVA queries, while the original PVA queries were replaced with the modified queries, and the inference process is the same as the PVA~\cite{staab2024beyond} because the mentioned attack is the SOTA PVA methods.

For the metrics, we leveraged the attack success rate (ASR) of PVA to evaluate the defense performance for the PVA.
The ASR is a ratio between the number of correct inferring attributes $N_{attack}$ and the total number of attributes $N_{total}$, which is formulated as follows:

\begin{equation}
    ASR=\frac{N_{total}}{N_{attack}}
\end{equation}

Moreover, the total time cost for inferring 100 attack queries is used to evaluate the defense efficiency of the tested methods under the same hardware conditions.
Because the GCR-Defender is not a client-side but a provider-side defense method, the utility and fluency evaluation of the defended query yielded by the RCG-Defender was not evaluated.
Meanwhile, due to the assumption mentioned in Section 3.1, the proposed retrieval-confused generation does not impact the generation performance of the common queries.
\begin{figure*}[t]
\centering
\includegraphics[width=1.0\textwidth]{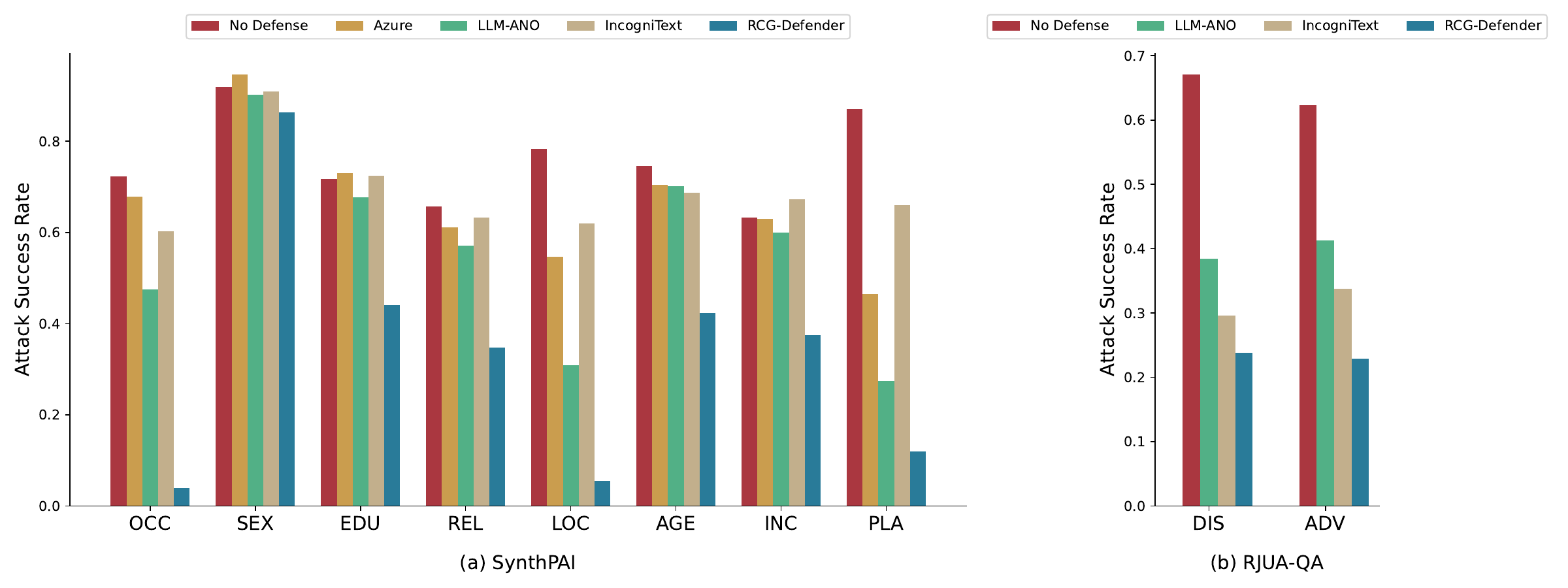}
\caption{The average attack success rate (ASR) of eight LLMs for different attributes in SynthPAI and RJUA-QA.}
\label{fig2:att-ASR}
\end{figure*}
\subsection{Main Results}
Since defense performance (i.e., attack success rate) and defense efficiency (namely, average time cost for defending each PVA query) are two key metrics for the privacy violation defense method, we evaluated the superiority of the proposed RCG-Defender method from the two aspects. Note that due to unsupported Chinese text anonymization, the Azure method is not suitable for the RJUA-QA dataset.
Thus, there are no corresponding experimental results in Table~\ref{tab:ASR} and Figure~\ref{fig2:att-ASR}.

\textbf{Defense performance.} We presented our main experiments in Table 1 and Figure~\ref{fig2:att-ASR}.
From Table~\ref{tab:ASR}, we compare the proposed RCG-Defender method with baselines in terms of the ASR on different mainstream LLMs, where the result values denote the average ASR for all attributes of two datasets.
We can draw the following conclusions:
(1) Compared with three SOTA baselines, the RCG-Defender achieves the best defense performance, i.e., the lowest average ASR. 
For instance, in the synthPAI, the RCG-Defender method achieves an average 26.27\% and 37.02\% ASR decrease compared with LLM-ANO and IncogniText, respectively.
(2) Obviously, the performance improvement on the synthPAI dataset is better than that on the RUJA-QA data.
For example, compared with LLM-ANO, the RCG-Defender gains 26.27\% and 17.14\% improvement, respectively.
The main reason is that most LLMs are good at inferring the English query. 
(3) Particularly for the GLM and Q72, the defense performance on RUJA-QA is better than that on SynthPAI since the two LLMs are training on more Chinese data compared to other tested LLMs. 
For the GLM, the ASR is lower than 30.14\% and 47.32\% on synthPAI and RUJA-QA, respectively.

As shown in Figure~\ref{fig2:att-ASR}, we compared defense performance on the \textit{SynthPAI} dataset with eight private attributes and the \textit{RJUA-QA} dataset with two private attributes, where each bin denotes an average ASR value of the tested defense method on eight mainstream LLMs. 
From the experimental results, we can draw the following conclusions:
(1) The proposed RCG-Defender method significantly outperforms other tested defense methods in terms of defense performance on StmthPAI and RJUA-QA datasets.
(2) For the ``SEX'' attribute of the SynthPAI dataset, the ASRs of the tested method still have high values (all results are over 80\%) because the corresponding text contents of user comments are highly explicit, leading to easier inference than other attributes.

\begin{figure}[t]
\centering
\includegraphics[width=\columnwidth]{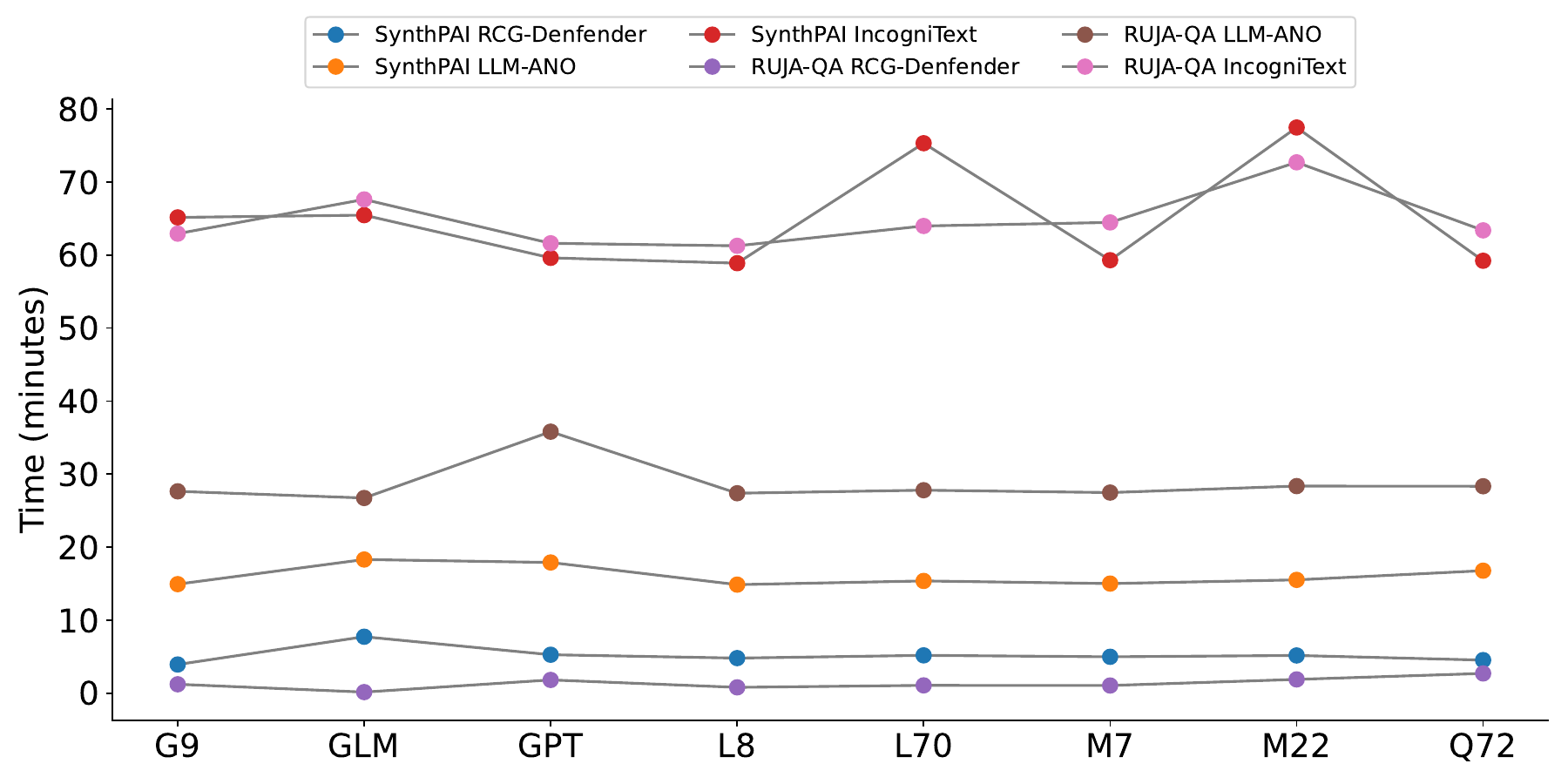}
\caption{The time cost of the RCG-Denfender and LLM-ANO inferring 100 attack queries on two datasets and eight LLMs.}
\label{fig:inference-time}
\end{figure}

\textbf{Defense efficiency.} 
In this part, we calculated the average time cost for defending 100 PVA queries on three tested methods, including LLM-ANO, IncogniText, and RCG-Defender.
Looking at the experimental results in Figure~\ref{fig:inference-time}, we observe that the RCG-Defender method is superior to the LLM-ANO and IncogniText.
In particular, for the proposed RCG-Defender, the experimental results on different LLMs and datasets are under 10 seconds.
The main reason is that the LLM-ANO and IncogniText methods must iteratively modify each input query until it meets the stop requirement, while the RCG-Defender merely makes one retrieval operation and one LLM inference operation for each PVA query.

\subsection{Ablation Study}
As stated in Section 4, the proposed RCG-Defender method has three important components: the disturbed database, the retrieval strategy, and the embedding model.
In this part, some experiments were carried out to evaluate the effectiveness of the proposed RCG-Defender in terms of the above three components.
Specifically, we conducted comprehensive ablation studies on the \textit{synthPAI} dataset and four LLMs: Llama-3-8B (L8), Llama-3-70B (L70), Qwen2.5-72B (Q72), and Gemma-9B (G9).

\begin{figure}[t]
\centering
\includegraphics[width=\columnwidth]{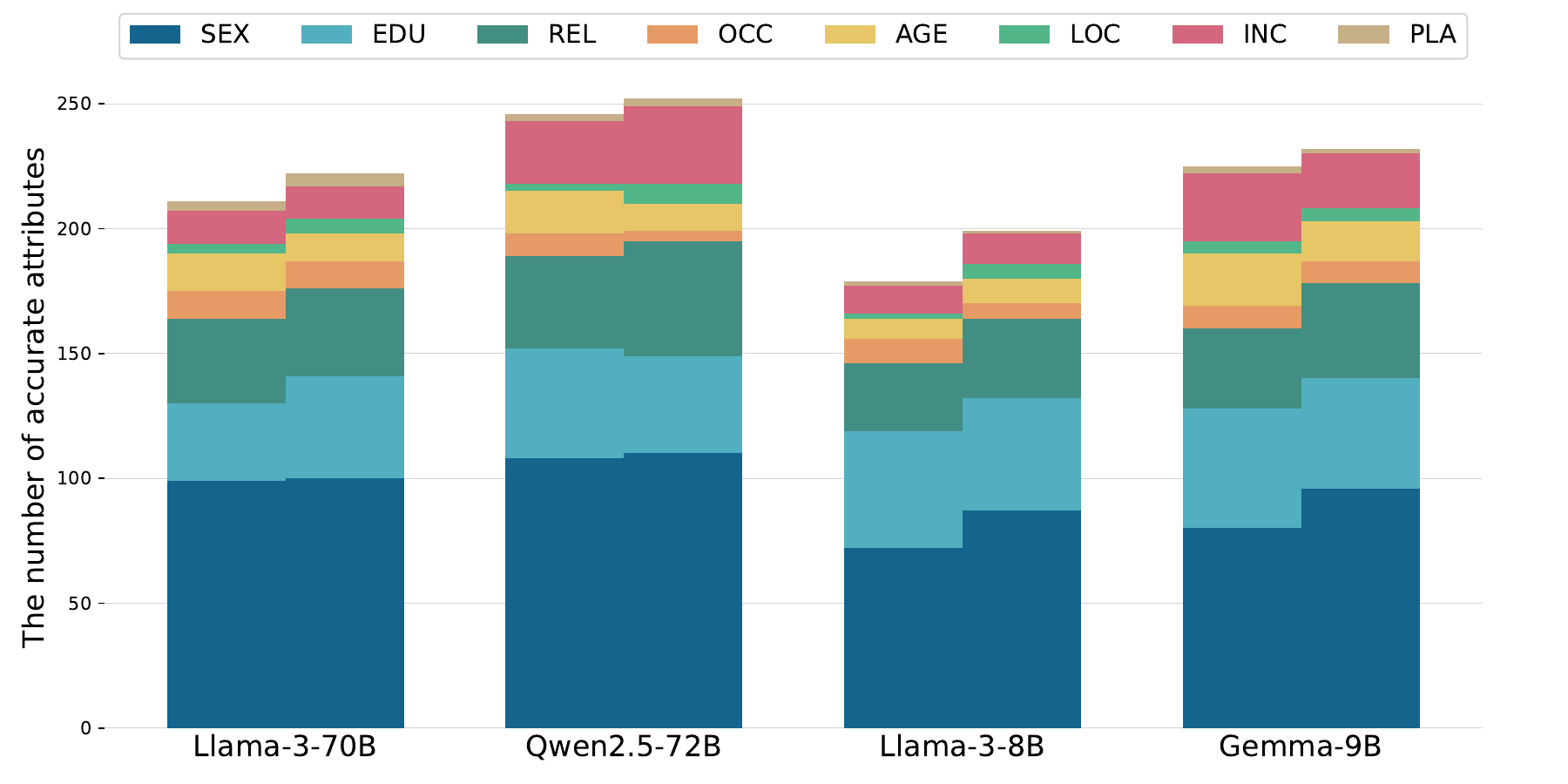}
\caption{The ablation experimental results for the disturbed database (the left side denotes the LLM-based database; the right side denotes the original database).}
\label{fig:database}
\end{figure}

\textbf{(1) Disturbed Database.}
In this part, we evaluated the effectiveness of the disturbed database constructed by the LLM with a paraphrasing prompt.
We constructed a control group, which contains the original comments without being paraphrased by the LLMs.
As shown in Figure~\ref{fig:database}, it can be observed that the disturbed database constructed by the paraphrasing prompt and LLM significantly improves the defense performance.
In particular, there are significant enhancements to the Llama series. 
Moreover, for different attributes and LLMs, the LLM-based database has side effects, such as the education and age attributes for Qwen2.5-72B.
The main reason is that the paraphrasing prompt is a manual design, which constrains the generalization for different LLMs.
This will be a critical issue explored in the future work.
In addition, the distrubed dataset can effectively avoid the leaking risk of the original user's comments, enhancing the data security of the confused-retrieval operation in the practical scenario.
\begin{table*}[t]
\renewcommand{\arraystretch}{1.3}
\centering
\caption{The ablation experimental results for the embedding model. (The bold and the underline indicate the best and sub-optimal results, respectively.)}
\label{tab:embed_model}
\begin{tabular}{c|c|ccccccccc}
\toprule[1pt]
\textbf{LLMs} & \textbf{Embeddings} & \textbf{AGE} & \textbf{EDU} & \textbf{INC} & \textbf{LOC} & \textbf{OCC} & \textbf{PLA} & \textbf{REL} & \textbf{SEX} & \textbf{Average}
\\ \hline
\multirow{3}{*}{G9} & all-Mini & 0.6111 & 0.3400 & 0.4222 & 0.0875 & 0.0386 & 0.0400 & 0.3125 & 0.7748 & \textbf{0.2957} \\ 

 & e5-base & 0.6111 & 0.3800 & 0.6000 & 0.0750 & 0.0483 & 0.1600 & 0.3958 & 0.8739 & 0.3457 \\ 
 
 & bge-large & 0.5833 & 0.4800 & 0.6000 & 0.0625 & 0.0435 & 0.1200 & 0.3333 & 0.7207 & \underline{0.3214} \\ 
 \hline
 
\multirow{3}{*}{L8} & all-Mini & 0.4444 & 0.4800 & 0.3556 & 0.0625 & 0.0386 & 0.0000 & 0.2917 & 0.3937 & 0.3229 \\ 

     & e5-base & 0.3333 & 0.4300 & 0.4889 & 0.1125 & 0.0338 & 0.1200 & 0.2708 & 0.7658 & \underline{0.3214} \\ 
 
 & bge-large & 0.2222 & 0.4700 & 0.2444 & 0.0250 & 0.0483 & 0.0800 & 0.2813 & 0.6486 & \textbf{0.3014} \\ 
 \hline
 
\multirow{3}{*}{L70} & all-Mini & 0.5000 & 0.3300 & 0.4000 & 0.0875 & 0.0531 & 0.0400 & 0.3333 & 0.9550 & \underline{0.2829} \\ 

 & e5-base & 0.4167 & 0.3400 & 0.2889 & 0.0500 & 0.3865 & 0.0800 & 0.4375 & 0.9640 & 0.2957 \\ 
 
 & bge-large & 0.4167 & 0.3100 & 0.2889 & 0.0500 & 0.0531 & 0.1600 & 0.3542 & 0.8919 & \textbf{0.2557} \\ 
 \hline

\multirow{3}{*}{Q72} & all-Mini & 0.4722 & 0.3900 & 0.5556 & 0.0750 & 0.0483 & 0.0000 & 0.3646 & 0.9820 & \textbf{0.3443} \\ 

 & e5-base & 0.3611 & 0.3700 & 0.5778 & 0.0250 & 0.0338 & 0.0800 & 0.5000 & 0.9910 & \underline{0.3500} \\ 
 
 & bge-large & 0.4722 & 0.4400 & 0.5556 & 0.0375 & 0.0435 & 0.1200 & 0.2854 & 0.9730 & 0.3514 \\ 
 \bottomrule[1pt]
\end{tabular}
\end{table*}

\begin{figure}[t]
\centering
\includegraphics[width=\columnwidth]{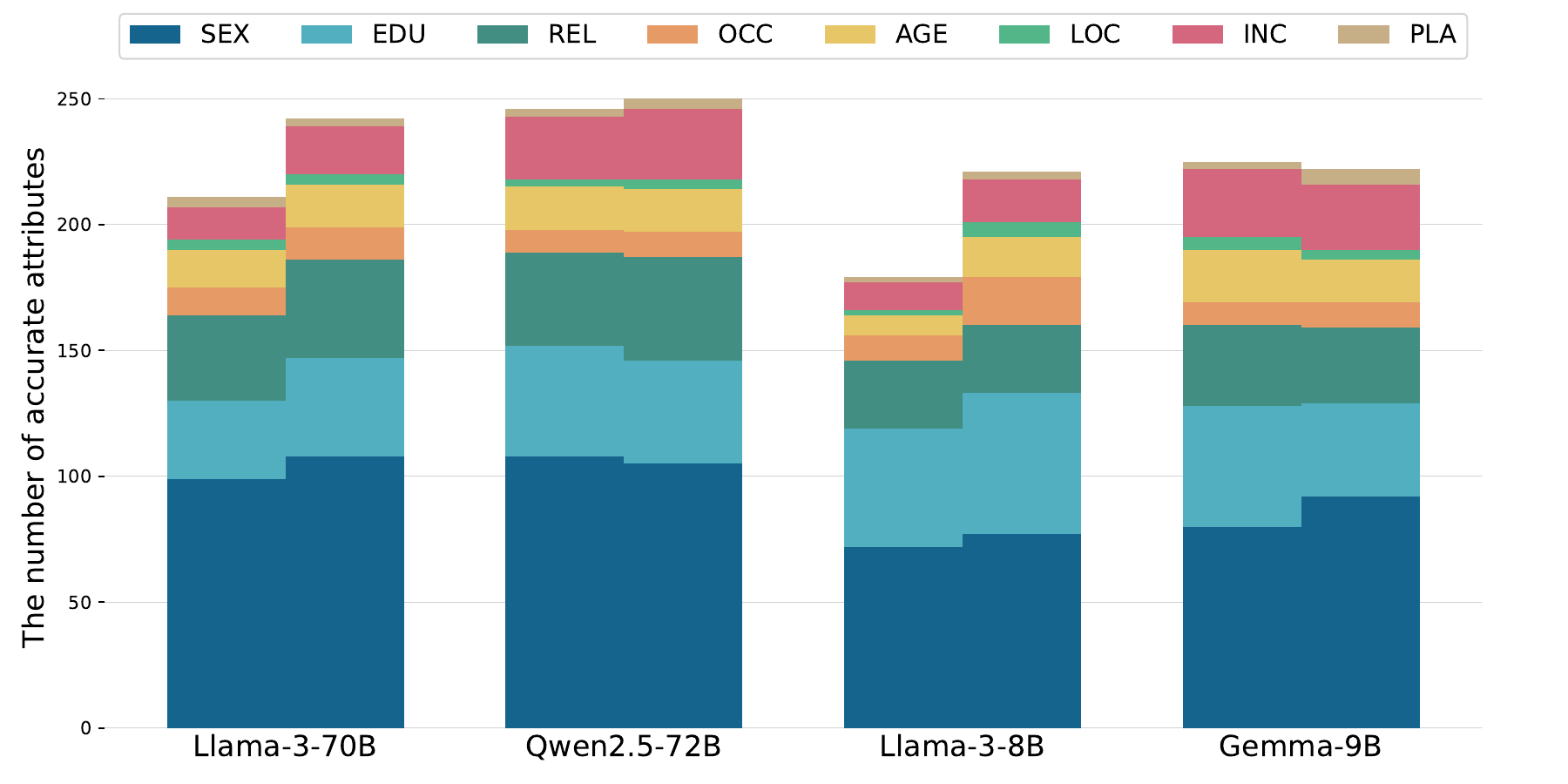}
\caption{The ablation experimental results for the retrieval strategy (the left side denotes the most irrelevant strategy; the right side denotes the random strategy).}
\label{fig:retrieval}
\end{figure}

\textbf{(2) Retrieval Strategy.} To examine the impact of different retrieval strategies on the defense performance of the RCG-Defender, 
we compared the most irrelevant and random retrieval strategies.
Note that the random retrieval strategy is that the retrieved comment is randomly selected from the disturbed database constructed by the LLM and paraphrasing prompt.
The experimental results, shown in Table~\ref{fig:retrieval}, indicate that the most irrelevant retrieval strategy shows better performance than that of the random one. 
For instance, the most irrelevant retrieval strategy gains the remarkable performance enhancement on the Llama3-8B.
Obviously, for the Gemma-9B model, there is an abnormal result, i.e., the proposed retrieval strategy has a side effect on the EDU and REL attributes.
The experimental results demonstrate that the generalization of the proposed retrieval strategy still has room for improvement.

\textbf{(3) Embedding Model.}
In this part, we presented an additional ablation study on the impact of three different embedding models of the proposed RCG-Defender method, including all-MiniLM-L6-v2 (all-mini), e5-based-v2 (e5-base), and bge-large-en-v1.5 (beg-large).
As shown in Table~\ref{tab:embed_model}, the defense performance varies significantly across different embedding models.
From the experimental results, we can draw the following conclusions:
(1) There is no general embedding model to achieve the best performance on different LLMs.
For the Llama series and Gemma-9B, the bge-large-en-v1.5 and all-MiniLM-L6-v2 are the best suitable embedding models, respectively.
(2) Compared with other models, the bge-large-en-v1.5 can achieve the comprehensively best performance for the PVA attack, and the e5-based-v2 model gains the worst ones.
\section{Conclusion}
Recent advances in LLMs have made a profound impact on our society and also raised new security concerns. 
Particularly, the privacy violation attack (PVA) introduces serious personal privacy issues.
In this paper, we presented a provider-side defense method via retrieval-confused generation (RCG-Defender), which can effectively and covertly defend against the emerging privacy violation attack (PVA) of LLM inference.
The RCG-defender contains two key stages: the disturbed database construction and the confused retrieval.
In the first stage, we design a paragraphing prompt to induce LLMs to rewrite the attack query content associated with personal attributes, so as to construct the disturbed database.
In the confused retrieval stage, we proposed the most irrelevant retrieval strategy to retrieve satisfied comments from the disturbed database.
Meanwhile, the ``user comments'' of the attack query are replaced with the retrieved comments, making the LLMs generate some wrong personal attributes. 
We carry out extensive experiments on two datasets and eight downstream LLMs, which demonstrates the superiority of the proposed method against the PVA.
The RCG-Defender is the first exploration of the provider-side defense method based on retrieval mechanisms.
We hope this attempt will open a new defense paradigm for the PVA of LLMs.

\bibliographystyle{ACM-Reference-Format}
\bibliography{custom}

\appendix









\end{document}